# Bounding Bias Due to Selection

*Louisa H. Smith[a] and Tyler J. VanderWeele[a,b]*

**Abstract:** When epidemiologic studies are conducted in a subset of the population, selection bias can threaten the validity of causal inference. This bias can occur whether or not that selected population is the target population and can occur even in the absence of exposure–outcome confounding. However, it is often difficult to quantify the extent of selection bias, and sensitivity analysis can be challenging to undertake and to understand. In this article, we demonstrate that the magnitude of the bias due to selection can be bounded by simple expressions defined by parameters characterizing the relationships between unmeasured factor(s) responsible for the bias and the measured variables. No functional form assumptions are necessary about those unmeasured factors. Using knowledge about the selection mechanism, researchers can account for the possible extent of selection bias by specifying the size of the parameters in the bounds. We also show that the bounds, which differ depending on the target population, result in summary measures that can be used to calculate the minimum magnitude of the parameters required to shift a risk ratio to the null. The summary measure can be used to determine the overall strength of selection that would be necessary to explain away a result. We then show that the bounds and summary measures can be simplified in certain contexts or with certain assumptions. Using examples with varying selection mechanisms, we also demonstrate how researchers can implement these simple sensitivity analyses. See video abstract at, http://links.lww.com/EDE/B535.

(*Epidemiology* 2019;30: 509–516)

**Editor's Note:** A related commentary appears on p. 517.

Submitted August 4, 2018; accepted April 8, 2019.
From the [a]Department of Epidemiology, Harvard T.H. Chan School of Public Health, Boston, MA; and [b]Department of Biostatistics, Harvard T.H. Chan School of Public Health, Boston, MA.
Supported by grant R01CA222147 from the NIH (T.V.).
Disclosure: The authors report no conflicts of interest.
Data and computing code: The R package EValue contains functions to implement these methods.
Supplemental digital content is available through direct URL citations in the HTML and PDF versions of this article (www.epidem.com). This content is not peer-reviewed or copy-edited; it is the sole responsibility of the authors.
Correspondence: Louisa H. Smith, Department of Epidemiology, Harvard T.H. Chan School of Public Health, 677 Huntington Avenue, Boston, MA 02115. E-mail: louisa_h_smith@g.harvard.edu.

When bias in an epidemiologic study is unavoidable, various methods can be used to assess the robustness of results to factors that limit causal inference, such as unmeasured confounding, measurement error, and selection. While there exist relatively simple sensitivity analysis approaches for measurement error and unmeasured confounding,[1–5] those for selection bias are limited by computational or mathematical complexity, the need for strong assumptions or the specification of a large number of parameters, or applicability only to certain selection mechanisms or study designs.[6–13]

In this article, we show that selection bias can be bounded by straightforward expressions that allow for a simple approach to sensitivity analysis. We use the term selection bias to describe the extent to which a parameter being estimated differs from a causal effect in either the total population or some subset of it, due to the restriction of the study population. This bias is sometimes defined as that resulting from selecting on a collider—that is, a common effect of two variables in the causal structure.[14] Such selection may be due to convenience in study design or analysis, a desire to evaluate an exposure–outcome relationship in a subset of the population, an attempt to limit other types of bias, or nonparticipation and loss to follow-up.

For example, in birth defects studies, it is often difficult or impossible to collect data on pregnancies that do not result in a live birth. Analyzing the exposure–outcome relationship in only live births may lead to selection bias when a factor determining the probability of live birth is also related to the exposure of interest. In other types of studies, a question about a particular subpopulation is the motivation for the selection. For example, an exposure (e.g., obesity) may be particularly harmful or protective among people with certain health conditions (e.g., cardiovascular disease). However, even when selection is due to interest in a particular subpopulation, selecting on a factor of interest does not eliminate the potential for bias, if for example that factor itself is related to the exposure.

Solutions to these problems often involve a number of assumptions, particularly when the target of causal inference is the whole population and not just the subpopulation from which the sample was selected. Our simplified approach to assessing selection bias makes clear the target of inference and limits the number of parameters and assumptions that determine the possible magnitude of the bias. We show that the magnitude of the bias in the causal risk ratio can be bounded by simple expressions that relate the variables in the causal structure, which may be known or hypothetical. In the eTable and eAppendix (http://links.lww.com/EDE/B521), we extend the results to the risk difference scale. The bounds differ depending on the target population of interest and the selection procedure but require no assumptions about the type or number of measured or unmeasured variables that cause the bias, or interactions between pairs of variables. We consider several causal structures under which the bounds can be applied and motivate their use in the contexts above and with other examples. Finally, we show that under certain assumptions about the equality of the parameters determining bias, a summary measure can be constructed for each of various scenarios, which can be used as a simple technique for assessing the robustness to selection bias of results from an epidemiologic analysis.

## THE SELECTION BIAS BOUNDING FACTOR

### The Size and Structure of the Bias

Consider a situation in which a causal population-level risk ratio (RR) comparing two levels of an exposure denoted $A$ is the parameter of interest. Although our results hold comparing any two values of categorical or continuous $A$, we will assume binary $A \in \{0,1\}$ for ease of notation. Let $Y \in \{0,1\}$ denote the binary outcome and $S$ be a binary indicator of selection, where $S = 1$ indicates the subset of the population included in the study and $S = 0$ indicates the subset of the population excluded in the study. Let $C$ denote a set of measured covariates. In case–control studies, the odds ratio (OR) may approximate the RR; we assume that this approximation holds throughout this article. Furthermore, although cases are selected with higher probability than controls in such studies, we can ignore that aspect of the selection mechanism, as it does not bias the OR.

We will use potential outcome notation wherein $Y_a$ indicates the value of $Y$ under treatment $A = a$. Let the causal RR conditional on covariates $C = c$, $P(Y_1 = 1|c)/P(Y_0 = 1|c)$ be denoted $RR_{AY}^{true}$ and assume that it is identifiable as $P(Y = 1|A = 1, c)/P(Y = 1|A = 0, c)$. This requires that certain identifiability conditions hold, including consistency, positivity, and, in particular, exchangeability $Y_a \perp\!\!\!\perp A | C$; that is, that $Y_a$ is independent of actual exposure $A$ conditional on measured covariates $C$.[15] For simplicity in the development that follows, we will assume that all analyses are carried out within strata of measured confounders $C$ as necessary and exclude reference to those variables, but all subsequent probability expressions can be interpreted as conditional on measured covariates $C$.

Suppose now, due to some selection mechanism, we are limited to estimating the RR only within a subpopulation, denoted by $S = 1$, so that we estimate $RR_{AY}^{obs} = P(Y = 1|A = 1, S = 1)/P(Y = 1|A = 0, S = 1)$. If we restrict analysis to $S = 1$, selection bias occurs if it is not the case that $Y_a \perp\!\!\!\perp A | S = 1$, even though $Y_a \perp\!\!\!\perp A$ in the population.

The bias is not due to the fact that the RRs in the total and selected populations differ but to the fact that $RR_{AY}^{obs}$ is not a causal effect even in the selected population. (Later in the text, Results 5A and 5B correspond to the situation in which such an effect is of interest.)

Several causal structures in which selection bias may occur are shown in the Figure. In each situation, bias is induced by a selection process, which is itself differential with respect to the exposure or outcome and some unmeasured (or possibly measured) factor(s), denoted $U$. Consider the setting in which conditional on some unmeasured covariate(s) $U$, we have $Y \perp\!\!\!\perp S | \{A, U\}$. This independence holds in the causal diagrams in the Figure A, B, and D. For ease of notation, we will consider $U$ to be a categorical variable, but the results

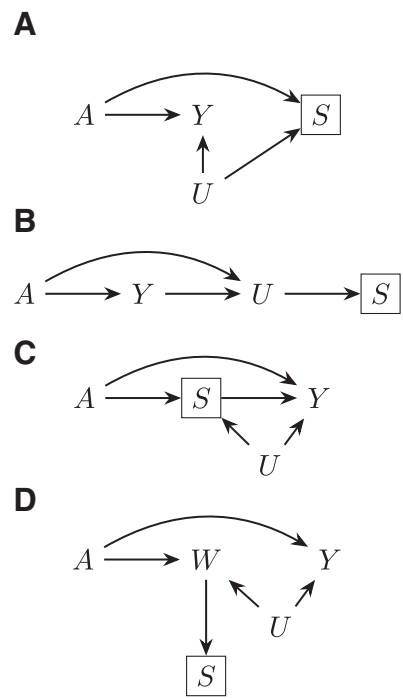

**FIGURE.** Examples of causal diagrams under which bias for a causal effect of $A$ on $Y$ can result due to selection. Selection is shown with the boxed $S$, representing that only a selected population, which may have different distributions of the variables upstream of $S$ than the total population, is studied. For simplicity, no confounders of the $A$–$Y$ relationship are shown but could certainly be present. A, The birth defects example, where $A$ is infection with Zika virus, $Y$ is microcephaly, $S$ is live birth, and $U$ is socioeconomic and behavioral factors. B, The estrogen cancer example, where selection is based on a factor $U$ (symptoms that lead to an intraendometrial diagnostic procedure) that is affected by both estrogen use ($A$) and the presence of endometrial cancer ($Y$). C, The obesity paradox example, where $A$ is obesity, $S$ is a disease such as heart failure, and $Y$ is mortality. D, A study in which only participants who agree to DNA collection ($S$) are selected to a study of the relationship between (1) a genetic risk score ($A$) for smoking ($W$) and (2) an education outcome ($Y$).[31]

hold for general $U$ or vector of variables denoted $U$. We can rewrite the target parameter in terms of $U$:

$$RR_{AY}^{true} = \frac{\sum_{s=0}^{1}\{\sum_u P(Y=1|A=1,S=s,U=u) P(U=u|A=1,S=s)\}P(S=s|A=1)}{\sum_{s=0}^{1}\{\sum_u P(Y=1|A=0,S=s,U=u) P(U=u|A=0,S=s)\}P(S=s|A=0)}.$$

Let the relative bias due to selection be defined as $RR_{AY}^{obs} / RR_{AY}^{true}$. By bounding this value, we can assess the maximum strength of the bias in terms of parameters that describe relationships between $U$ and other variables.

## The Bounding Factor

We bound the relative bias from above, assuming that the $RR_{AY}^{obs} > RR_{AY}^{true}$. If not, and the bias < 1, interest is naturally in a bound from below. We can then reverse the coding of $A$ so that bias > 1, resulting in an appropriate bound once the coding is reversed.

We define the following parameters:

$$RR_{UY|(A=1)} = \frac{\max_u P(Y=1|A=1,U=u)}{\min_u P(Y=1|A=1,U=u)},$$

$$RR_{UY|(A=0)} = \frac{\max_u P(Y=1|A=0,U=u)}{\min_u P(Y=1|A=0,U=u)},$$

$$RR_{SU|(A=1)} = \max_u \frac{P(U=u|A=1,S=1)}{P(U=u|A=1,S=0)},$$

$$RR_{SU|(A=0)} = \max_u \frac{P(U=u|A=0,S=0)}{P(U=u|A=0,S=1)}.$$

The $RR_{UY|(A=a)}$ parameters can be interpreted as the maximum relative risks for $Y = 1$ comparing any two values of $U$ within strata of $A = 1$ and $A = 0$, respectively. It need not be a causal relationship that is described by this risk ratio, as $U$ may be downstream of $Y$ in some situations that are susceptible to selection bias (e.g., Figure B). The $RR_{SU|(A=a)}$ parameters are the maximum factors by which selection is associated with an increased prevalence of some value of $U$ within stratum $A = 1$ and by which nonselection is associated with an increased prevalence of some value of $U$ within stratum $A = 0$. If $RR_{AY}^{obs}$ has been estimated within strata of measured confounders $C$, then these parameters are defined conditional on the same confounders.

We now present our first result, a proof of which is given in the eAppendix (http://links.lww.com/EDE/B521).

## Result 1A

If $Y \perp\!\!\!\perp S \mid \{A,U\}$, then:

$$\frac{RR_{AY}^{obs}}{RR_{AY}^{true}} \leq \left(\frac{RR_{UY|(A=1)} \times RR_{SU|(A=1)}}{RR_{UY|(A=1)} + RR_{SU|(A=1)} - 1}\right) \times \left(\frac{RR_{UY|(A=0)} \times RR_{SU|(A=0)}}{RR_{UY|(A=0)} + RR_{SU|(A=0)} - 1}\right).$$

Result 1A tells us that the bias is guaranteed to be equal to or smaller in magnitude than the given expression, which we will call in general a bounding factor. A researcher or reader who proposes that some factor $U$ has led to selection bias can propose values that plausibly describe the relationships between that factor and selection and the outcome and calculate the bounding factor from these parameters. That bounding factor (or set of bounding factors constructed from ranges of values) can be divided out of the estimate of $RR_{AY}^{obs}$ to come up with the smallest possible RR that would be compatible with $RR_{AY}^{true}$.

## Example: Zika Virus

After a rise in microcephaly cases in northeast Brazil closely followed an outbreak of Zika virus in that region, evidence from biologic and ecologic data supported a causal link.[16] In particular, models using surveillance data showed that the population risk of microcephaly increased after Zika infections in the first semester of pregnancy.[17] The relationship was seemingly confirmed with the first case–control study to examine the association, from which de Araújo et al reported an adjusted OR of 73.1 (95% confidence interval [CI] = 13.0, ∞).[18] Both live and still births were recruited as cases; however, pregnancies that resulted in miscarriage or elective abortion would have been missed by this study design, which corresponds to Figure A. The probability of not having a termination ($S = 1$) may be affected by exposure to the virus ($A$) and socioeconomic or behavioral conditions such as lack of access to medical care ($U$), which may also affect the probability of microcephaly ($Y$) (e.g., giving birth in a public hospital, low education, and being unmarried have been associated with microcephaly in Brazil[19]). Selecting only live and still births in the analysis may therefore lead to selection bias.

Suppose that access to medical care affected the probability of microcephaly by up to two-fold among the Zika exposed and unexposed (i.e., $RR_{UY|(A=1)} = RR_{UY|(A=0)} = 2$) and that lack of access to medical care for pregnant women was up to 1.7 times more likely for women without an induced abortion among the Zika exposed ($RR_{SU|(A=1)} = 1.7$) and access to medical care up to 1.5 times more likely for women with an induced abortion among the unexposed ($RR_{SU|(A=1)} = 1.5$), the bias factor is then

$$\left(\frac{RR_{UY|(A=1)} \times RR_{SU|(A=1)}}{RR_{UY|(A=1)} + RR_{SU|(A=1)} - 1}\right) \times \left(\frac{RR_{UY|(A=0)} \times RR_{SU|(A=0)}}{RR_{UY|(A=0)} + RR_{SU|(A=0)} - 1}\right)$$

$$= \left(\frac{2 \times 1.7}{2 + 1.7 - 1}\right) \times \left(\frac{2 \times 1.5}{2 + 1.5 - 1}\right) = 1.51.$$

The most such selection bias that could alter the estimate can be obtained by dividing the original estimate and confidence interval by this bias factor to obtain an OR of 48.4 (95% CI = 8.6, ∞), which is of course still a very large effect estimate.

### A Summary Measure

Instead of calculating each of the parameters in the bounding factor individually, we may be interested in assessing the overall susceptibility of a result to selection bias. This can be done with a single value that summarizes the extent to which an $RR_{AY}^{obs} > 1$ may be a spurious finding entirely due to selection bias.

### Result 1B

If $Y \perp\!\!\!\perp S \mid \{A, U\}$, then the minimum magnitude of each of the four parameters that make up the bounding factor, assuming the four are equal, that would be sufficient to shift a given $RR_{AY}^{obs}$ to the null is given by the following equation:

$$RR_{UY|(A=0)} = RR_{UY|(A=1)} = RR_{SU|(A=0)}$$
$$= RR_{SU|(A=1)} \geq \sqrt{RR_{AY}^{obs}} + \sqrt{RR_{AY}^{obs} - \sqrt{RR_{AY}^{obs}}}.$$

For example, if $RR_{AY}^{obs} = 3$, then all four of the parameters in the bounding factor must be equal to or greater than $\sqrt{3} + \sqrt{3 - \sqrt{3}} = 2.9$ to have generated sufficient selection bias. If one of the four is smaller than 2.9, then one or more of the parameters must be greater than 2.9 to compensate. Because it only depends on $RR_{AY}^{obs}$, this summary measure is easy to calculate and compare across studies. However, it is context specific: it is interpreted relative to the selection mechanism in a given study and conditional on whatever confounders have been controlled for in the analysis. Based on content knowledge, investigators and readers can judge whether there exists an $U$ that could be so strongly related both to the outcome and to selection within strata of the exposure and the measured confounders.

Such calculations can also be performed using the lower limit of the confidence interval instead of $RR_{AY}^{obs}$, to see what strength of the selection parameters would be necessary to result in a confidence interval that includes the null value of 1.

### Example: Zika Virus Revisited

With no assumptions about the exact nature of the unmeasured factors $U$, we can use Result 1B to assess the plausibility that the Zika–microcephaly association is fully explained by selection bias. By calculating the summary measure $\sqrt{73.1 - \sqrt{73.1}} + \sqrt{73.1} = 16.6$, we come to conclusions about the strength of the relationships with the unmeasured behaviors or socioeconomic conditions (such as lack of access to medical care) that would be necessary to produce an $RR_{AY}^{obs}$ of 73.1 if $RR_{AY}^{true} = 1$. Selection bias could explain the observed association if there was an unmeasured variable (e.g., lack of access to medical care) that increased the risk of microcephaly by 16.6-fold in both exposed and unexposed women, that was 16.6 times higher among exposed women with live or still births than among those whose pregnancies were terminated, and that was also 16.6 times lower among the unexposed. However, weaker relationships between the unmeasured factor and both selection and microcephaly would not suffice to fully explain the observed exposure–outcome association. Risk ratios of that magnitude are rarely seen in epidemiologic research, particularly in the context of behavioral differences, lending confidence that the increased microcephaly risk is not the result of selection bias. We can repeat the calculation with the lower bound of the confidence interval, 13.0, to assess the magnitude of selection bias necessary to shift the confidence interval to include the null. This gives a summary measure of 6.7; although it is perhaps plausible that one of the parameters is that large, it seems unlikely that all four are. Assuming that confounding was fully accounted for in the study via matching and multivariable control and that all variables were correctly measured, it seems that even in the presence of possible selection bias, the evidence is very strong that Zika infection in pregnant women causes microcephaly.

## SPECIAL CASES AND THEIR BOUNDING FACTORS AND SUMMARY MEASURES

Here, we consider a number of special cases that result in modified bounding factors and summary measures. The Table summarizes the results, and derivations are provided in the eAppendix (http://links.lww.com/EDE/B521).

### When $S = U$ (Results 2A and 2B)

In some situations, $U$ may not be unmeasured and may be common to the entire selected population. This is the case, for example, when some characteristic defines or directly leads to selection into a study. When this is true, the bounding factor is simplified.

### Result 2A

If $S = U$, then

$$\frac{RR_{AY}^{obs}}{RR_{AY}^{true}} \leq RR_{UY|(A=0)} \times RR_{UY|(A=1)}.$$

We can also construct a summary measure for this situation. It can be used in the same way as that in Result 1B but only describes the minimum magnitude of the two parameters in the modified bounding factor in Result 2A.

### Result 2B

If $S = U$, then the minimum magnitude of each of the two parameters that make up the bounding factor in Result 2A, assuming they are equal, that would be sufficient to shift a given $RR_{AY}^{obs}$ to the null is given by the following equation:

$$RR_{UY|(A=0)} = RR_{UY|(A=1)} \geq \sqrt{RR_{AY}^{obs}}.$$

TABLE. Summary of Bounding Factors for Selection Bias on the Risk Ratio Scale and Their Summary Measures Under Different Scenarios

| | Bounding Factor[a] (A) | Summary Measure[b] (B) |
|---|---|---|
| Result 1. General selection bias[c,d] | $\left(\dfrac{RR_{UY|(A=1)} \times RR_{SU|(A=1)}}{RR_{UY|(A=1)} + RR_{SU|(A=1)} - 1}\right) \times \left(\dfrac{RR_{UY|(A=0)} \times RR_{SU|(A=0)}}{RR_{UY|(A=0)} + RR_{SU|(A=0)} - 1}\right)$ | $\sqrt{RR_{AY}^{obs}} + \sqrt{RR_{AY}^{obs} - \sqrt{RR_{AY}^{obs}}}$ |
| Result 2. When $S = U$ [c,e] | $RR_{UY|(A=0)} \times RR_{UY|(A=1)}$ | $\sqrt{RR_{AY}^{obs}}$ |
| Result 3. Increased risk with selection in both exposure groups[c,f] | $\dfrac{RR_{UY|(A=1)} \times RR_{SU|(A=1)}}{RR_{UY|(A=1)} + RR_{SU|(A=1)} - 1}$ | $RR_{AY}^{obs} + \sqrt{RR_{AY}^{obs}(RR_{AY}^{obs} - 1)}$ |
| Result 4. $S = U$ and increased risk[c] | $RR_{UY|(A=1)}$ | $RR_{AY}^{obs}$ |
| Result 5. Inference in the selected population[g] | $\dfrac{RR_{UY|(S=1)} \times RR_{AU|(S=1)}}{RR_{UY|(S=1)} + RR_{AU|(S=1)} - 1}$ | $RR_{AY}^{obs} + \sqrt{RR_{AY}^{obs}(RR_{AY}^{obs} - 1)}$ |

[a]The relative bias due to selection of the observed risk ratio, $RR_{AY}^{obs} / RR_{AY}^{true}$, is guaranteed to be less than this value. The parameters that define each bounding factor are defined in the main text.
[b]If all of the parameters in the bounding factor are equal, then each must be greater than this value to shift $RR_{AY}^{obs}$ to 1.
[c]The bounding factor and summary measure hold under the assumption that $Y \coprod S | \{A,U\}$.
[d]The parameter of interest, $RR_{AY}^{true}$, is the causal risk ratio for the whole population.
[e]The factor responsible for selection bias, $U$, is common to the entire selected population.
[f] $P(Y=1|A=1, S=1) / P(Y=1|A=1, S=0)$ and $P(Y=1|A=0, S=1) / P(Y=1|A=0, S=0)$ are greater than 1.
[g]The parameter of interest, $RR_{AY|(S=1)}^{true}$, is the causal risk ratio in the selected population only. The bounding factor and summary measure hold under the assumption that $Y_a \coprod A | \{S=1, U\}$.

## Assumptions About Directionality (Results 3A and 3B)

Although Result 1A requires minimal assumptions, sometimes we can make assumptions that decrease the magnitude of the bounding factor, which can provide us with more confidence that a given result is not due to selection bias. The bounding factor is greatest when $P(Y=1|A=1, S=1) > P(Y=1|A=1, S=0)$ and $P(Y=1|A=0, S=1) < P(Y=1|A=0, S=0)$; that is, when selection is associated with increased risk of the outcome among the exposed and with decreased risk among the unexposed. However, if selection is associated with increased risk among both groups, then we have the following result.

### Result 3A

If $Y \coprod S | \{A,U\}$ and if $P(Y=1|A=1, S=1) / P(Y=1|A=1, S=0)$ and $P(Y=1|A=0, S=1) / P(Y=1|A=0, S=0)$ are greater than 1, then

$$\frac{RR_{AY}^{obs}}{RR_{AY}^{true}} \leq \frac{RR_{UY|(A=1)} \times RR_{SU|(A=1)}}{RR_{UY|(A=1)} + RR_{SU|(A=1)} - 1}.$$

Results are analogous with decreased risk for both groups, with $A = 0$ replacing $A = 1$ in each of the parameters. (If selection is associated with decreased risk among the exposed and increased risk among the unexposed, the bias $\leq 1$, so $A$ should be recoded to construct a meaningful bound.)

If assumptions about the consistency of the direction of the selection–outcome relationship can be made, then we can also use simpler expressions as the summary measures; for increased risk in both groups, the summary measure is stated in the following result.

### Result 3B

If $Y \coprod S | \{A,U\}$ and if $P(Y=1|A=1, S=1) / P(Y=1|A=1, S=0)$ and $P(Y=1|A=0, S=1)/P(Y=1|A=0, S=0)$ are greater than 1, then the minimum magnitude of each of the two parameters that make up the bounding factor in Result 3A, assuming they are equal, that would be sufficient to shift a given $RR_{AY}^{obs}$ to the null is given by the following equation:

$$RR_{UY|(A=1)} = RR_{SU|(A=1)} \geq RR_{AY}^{obs} + \sqrt{RR_{AY}^{obs}(RR_{AY}^{obs} - 1)}.$$

When the outcome risk is decreased with selection in both exposure groups, the summary measure refers to the minimum strength of the parameters $RR_{UY|(A=0)}$ and $RR_{SU|(A=0)}$. Results 3A and 3B have the same analytic form of the recently proposed "E-value" calculated to assess robustness to unmeasured confounding.[20]

## When $S = U$ with Directionality Assumptions (Results 4A and 4B)

When $S = U$ and we can make assumptions about the increase or decrease in risk in both exposure groups with selection, we can combine earlier results. For increased risk in both groups, we have the following result.

### Result 4A

If $S = U$ and if $P(Y=1|A=1, S=1) / P(Y=1|A=1, S=0)$ and $P(Y=1|A=0, S=1) / P(Y=1|A=0, S=0)$ are greater than 1, then

$$\frac{RR_{AY}^{obs}}{RR_{AY}^{true}} \leq RR_{UY|(A=1)}.$$

The summary measure describing the minimum magnitude of the sole parameter is also simplified.

### Result 4B

If $S = U$ and if $P(Y=1|A=1, S=1)/P(Y=1|A=1, S=0)$ and $P(Y=1|A=0, S=1)/P(Y=1|A=0, S=0)$ are greater than 1, then the minimum magnitude of $RR_{UY|(A=1)}$ that would be sufficient to shift a given $RR_{AY}^{obs}$ to the null is given by the following equation:

$$RR_{UY|(A=1)} \geq RR_{AY}^{obs}.$$

### Example: Endometrial Cancer

For many years, the relationship between estrogen replacement therapy and endometrial cancer was clouded by controversy over proper study design to minimize bias. In an attempt to limit differential outcome detection by estrogen exposure, Horwitz and Feinstein[21] simultaneously performed a case–control study of exogenous estrogens and endometrial cancer in a population of women who had undergone intraendometrial diagnostic procedures and one with a more "conventional" sampling method. They claimed their estimates of an OR of 2.30 using the "alternative" sampling method and an OR of 11.98 with the conventional method supported their worry about biased cancer detection. However, their selection procedure was shown to induce bias.[22] The structure leading to this bias is shown in Figure B, in which $U$ represents a diagnostic procedure; in this case, all of those selected ($S$) have undergone such a procedure. We can use a bounding factor to assess how plausible it is that selection bias could explain the authors' much reduced OR. In this context, we are curious about whether bias could shift the result to a specific value and not to the null.

Because our proposed $RR_{AY}^{true} = 11.98 > RR_{AY}^{obs} = 2.30$, the bias < 1 and we recode the exposure for a relative bias of $11.98/2.3 = 5.2$. Because everyone in the selected population had symptoms that led to a diagnostic procedure, we will use the bounding factor that assumes $S = U$. If we assume that having a hysterectomy is associated with increased cancer prevalence in both estrogen-exposed and unexposed women, we have by Result 4B that $5.2 < RR_{UY|(A=1)}$ (recalling that $A = 1$, after recoding, now refers to those *unexposed* to estrogen). This means that in order for the difference in ORs to be possibly explained by selection bias (that is, for the $RR_{AY}^{obs}$ of 2.30 to shift to at least 11.98 after accounting for selection bias), the prevalence of endometrial cancer in nonusers of estrogen who have undergone hysterectomy or other diagnostic procedure must be greater than 5.2 times that in nonusers who have not.

## THE SELECTED POPULATION AS THE TARGET POPULATION

In some studies, the causal risk ratio in the selected population, and not in the entire population, may be the target parameter. This may occur, for example, when $S$ is an indicator of a well-defined population for which an estimated causal effect is desired and not simply the result of poor sampling or selective attrition.

Under the same notation and assumptions as above, we denote $RR_{AY|(S=1)}^{true} = P(Y_1 = 1|S=1)/P(Y_0 = 1|S=1)$. Again, because it is not true that $Y_a \perp\!\!\!\perp A | S = 1$, this is not identifiable as $RR_{AY}^{obs} = P(Y=1|A=1, S=1)/P(Y=1|A=0, S=1)$. Instead, if $Y_a \perp\!\!\!\perp A | \{S=1, U\}$, the RR of interest is identified by marginalizing over the distribution of $U$ in the selected population, resulting in

$$RR_{AY|(S=1)}^{true} = \frac{\sum_u P(Y=1|A=1, S=1, U=u)P(U=u|S=1)}{\sum_u P(Y=1|A=0, S=1, U=u)P(U=u|S=1)}.$$

Again we are concerned with the relative bias $RR_{AY}^{obs}/RR_{AY|(S=1)}^{true}$. This bias can be conceptualized as equivalent to that due to unmeasured confounding, which occurs when $U$ is associated with the exposure $A$ and also affects the outcome $Y$. Although $A$ and $U$ are marginally independent, as in Figure A, C, and D, an association between the two is induced by conditioning on selection into the study, represented by the boxed $S$ in the diagrams. Then, within stratum $S=1$, we have a situation equivalent to confounding by $U$, due to its relationships with the exposure and outcome. Extending previously published bounds for bias due to unmeasured confounding,[5] we have a bounding factor for inference in the selected population as follows.

### Result 5A

If $Y_a \perp\!\!\!\perp A | \{S=1, U\}$, then

$$\frac{RR_{AY}^{obs}}{RR_{AY|(S=1)}^{true}} \leq \frac{RR_{UY|(S=1)} \times RR_{AU|(S=1)}}{RR_{UY|(S=1)} + RR_{AU|(S=1)} - 1},$$

where

$$RR_{UY|(S=1)} = \max_a \frac{\max_u P(Y=1|A=a, S=1, u)}{\min_u P(Y=1|A=a, S=1, u)},$$

$$RR_{AU|(S=1)} = \frac{\max_u P(U=u|A=1, S=1)}{\min_u P(U=u|A=0, S=1)}.$$

The parameter $RR_{UY|(S=1)}$ is the maximum risk ratio for the outcome given any two values of $U$ among either the unexposed selected population or the exposed selected population. Because data are available on a sample of this population, this could be approximated using available data on measured confounders.

Because the second parameter, $\text{RR}_{AU|(S=1)}$, represents an association induced between two marginally independent variables (i.e., the dependence due to collider stratification), it is not as intuitive to specify. However, it is conceptually similar to one of the two required to define a bound for bias in the natural direct effect,[23] where the $A$–$U$ relationship is induced by conditioning on a mediator. In that situation, an approximate bound can be used.[24] Similarly, the bound that uses the maximum RR for $S=1$ comparing two values of $U$ or the maximum risk ratio for $S=1$ comparing two values of $A$ could be used here. Depending on the structure of the selection bias mechanism, one of these two parameters might be more intuitive to quantify and can generally replace $\text{RR}_{AU|(S=1)}$ for an approximate bounding factor.

### Summary Measure

The summary measure that follows from Result 5A can then be given in the following result.

### Result 5B

If $Y_a \perp\!\!\!\perp A \mid \{S=1, U\}$, then the minimum value of $\text{RR}_{AU|(S=1)}$ and $\text{RR}_{UY|(S=1)}$, assuming the two parameters are equal, that would be sufficient to shift a given $\text{RR}_{AY}^{\text{obs}}$ to the null is given by the following equation:

$$\text{RR}_{UY|(S=1)} = \text{RR}_{AU|(S=1)} \geq \text{RR}_{AY}^{\text{obs}} + \sqrt{\text{RR}_{AY}^{\text{obs}}(\text{RR}_{AY}^{\text{obs}} - 1)}.$$

Results 5A and 5B have the same analytic form of the recently proposed E-value calculated to assess robustness to unmeasured confounding.[20]

### Example: Obesity Paradox

The obesity paradox is a well-known phenomenon in chronic disease epidemiology in which overweight and obesity are associated with increased survival compared with normal weight among patients with certain conditions.[25] Whether this is a real causal effect (which could result in different weight recommendations for people living with chronic conditions) or due to bias—in particular, bias resulting from selection on a common effect (chronic disease) of both obesity and some unmeasured factor that is also related to death (Figure C)—is the subject of much debate.[26–29] Gruberg et al[30] investigated the relationship between body mass index and 1-year risk of death among patients who were treated for advanced coronary artery disease (CAD), finding that 10.6% of patients with normal body mass index died, more than double the percentage among the obese patients. Using the OR from their adjusted model, we can calculate that the mortality risk was 1.50 times higher (95% CI = 1.22, 1.86) in patients for whom body mass index was 10 units lower (corresponding approximately to the difference between obesity and normal weight).[30] The authors controlled for a number of measured confounders including age and heart function.

Because we are interested in the population of CAD patients, we can use Result 5B to assess the plausibility of such a result being due to some unmeasured common cause of heart failure and death: $1.50 + \sqrt{1.50(1.50-1)} = 2.37$. The unmeasured factor must increase the risk of death among normal weight or obese CAD patients by a factor of 2.37 (independent of the factors already included in the model) and differ between the obesity exposure categories in CAD patients by the same factor, if there were truly no protective effect of obesity on death in that population. Because the latter relationship, between obesity and the unmeasured factor, would be one induced solely by the selection of CAD patients, it may be difficult to specify. It may be more intuitive to consider an unmeasured factor that directly increases the risk of CAD by the same factor of 2.37, which will generally also suffice to bound the selection bias. We can repeat the calculation and interpretation with the lower bound of the confidence interval, which gives us a summary measure of 1.74, to assess the bias necessary for the confidence interval to include the null.

## DISCUSSION

Because selection bias can be difficult to quantify, it is often ignored in sensitivity analysis or only explored in complex analyses that must be relegated to appendices. A simple way to characterize the possible extent of selection bias in terms of the relationships in the causal structure that induces it will allow researchers to more easily assess the plausibility of this bias with minimal assumptions.

Thinking about selection bias as described in this article will also force researchers to clearly define the target population of interest, whether that be the total population or those with the characteristics of the selected sample. Making assumptions to simplify the bounding factor can also compel them to think through the mechanisms by which selection bias occurs and the direction of the various effects. However, no such assumptions are required to use our main results. While this article focused on the relative bias of observed RRs, as relative effect measures are common in epidemiology and RRs are often approximated by ORs and hazard ratios under certain assumption, analogous bounds for observed effects on the risk difference scale are presented in the eTable (http://links.lww.com/EDE/B521), and their corresponding derivations in the eAppendix (http://links.lww.com/EDE/B521).

The bounds we presented in this article can be used in several ways. If researchers have quantitative knowledge about factors influencing selection in their study, such as in a situation with loss to follow-up or participation in a sub-study, realistic RRs for unmeasured factors can be used as parameters in the bounds to explore to what extent these could affect $\text{RR}_{AY}^{\text{obs}}$. If only ranges of possible parameters are proposed, the bounds can be varied across those ranges in a table or figure to allow readers to consider the most plausible combinations. Finally, if all that is desired is a summary measure of the extent to which a result could be rendered null by selection bias, or shifted to any other proposed true value, the bounds can be

　　　　　　　　　　　　　　　　　　　　　　

used to describe the magnitude of the parameters that could result in such an observed value.

There are nonetheless several limitations to these bounds. First, they are only applicable under certain causal structures that lead to selection bias. The results here describe the maximum bias that could result from the parameters; the same parameters could also induce less bias. This conservative approach is useful when less is known about the selection mechanism and a simple exploration of the possible bias is desired. When more information is available, a more complex but precise method may be preferred.[7,8,10–12] Next, the $RR_{AU|(S=1)}$ parameter in the bound for the selected population is unintuitive and may be hard to specify even in the presence of solid knowledge about the selection mechanism; however, RRs relating the exposure or selection to the unobserved factor can usually be used in its place.[24] Finally, this article only addresses bias due to selection and assumes that other criteria for causal inference, such as control of exposure–outcome confounding and lack of measurement error, have been met. Future work could combine this approach to selection bias with other methods for bias analysis and could take into account the possibility that factors leading to selection bias could be sources of other types of bias.